\documentclass[aps,pra,twocolumn,showpacs,superscriptaddress]{revtex4-1}
\usepackage{amsmath} 
\usepackage{amssymb}
\usepackage{mathtools}
\usepackage{float}
\usepackage{graphicx,graphics,color}
\renewcommand{\vec}[1]{\boldsymbol{#1}}
\begin{document}
\title{Total Energy Dynamics and Asymptotics of the Momentum Distribution  \\
Following an Interaction Quench in a Two-component Fermi Gas}
\author{Chen-How Huang}
\affiliation{Department of Physics, National Tsing Hua University, Hsinchu 30013, Taiwan}
\author{Miguel A. Cazalilla}
\affiliation{Department of Physics, National Tsing Hua University, Hsinchu 30013, Taiwan}
\affiliation{National
Center for Theoretical Sciences (NCTS), Hsinchu 30013, Taiwan}
\affiliation{Donostia International Physics Center (DIPC), Manuel de
Lardizabal, 4. 20018, San Sebastian, Spain} 

\date{\today}
\begin{abstract}
The absence of a characteristic momentum scale in the pseudo-potential description of atomic interactions in ultracold (two-component Fermi) gases is known to lead to divergences in perturbation theory. Here we show that they also plague the calculation of the dynamics of the total energy following a quantum quench. A procedure to remove the divergences is  devised, which provides finite answers for the time-evolution of the total energy after a  quench in which the interaction strength is ramped up according to an arbitrary protocol. An important result of this analysis is the time evolution of the asymptotic tail of the momentum distribution (related to Tan's contact) to leading order in the scattering length. 
Explicit expressions for the  dynamics of the total energy and the contact for a linear interaction ramp are obtained, as a function of the interaction ramp time in the crossover from the sudden quench to the adiabatic limit are reported.  In sudden quench limit, the contact, following a rapid oscillation, reaches a stationary value which is different from the equilibrium one. In the adiabatic limit, the contact grows quadratically in time and later  saturates to its equilibrium value for the final value of the scattering length.
\end{abstract}
\maketitle
\section{Introduction}

The single-channel model~\cite{LHYpseudopotential} provides a compact, single-parameter description of  interactions in  ultracold gases. Within this model, interactions in a two-component Fermi gas are described by the following term in the  Hamiltonian:
\begin{equation}
\hat{V}=\frac{g}{2\Omega} 
\sum_{\vec{pkqr}}\sum_{\sigma\neq\alpha}c^{\dagger}_{\vec{p}\sigma}c^{\dagger}_{\vec{k}\alpha}c_{\vec{q}\alpha}
c_{\vec{r}\sigma} \delta_{\vec{p}+\vec{k},\vec{q}+\vec{r}},\label{eq:U}
\end{equation}
where $c_{\vec{p}\sigma}$ ($c^{\dag}_{\vec{p}\sigma}$) are fermion destruction (creation) operators obeying $\{c_{\vec{p}\sigma},c^{\dag}_{\vec{k}\alpha}\} = \delta_{\vec{pk}} \delta_{\sigma \alpha}$ ($\sigma,\alpha = \uparrow,\downarrow$) and $\{c_{\vec{p}\sigma},c_{\vec{k}\alpha}\} = 0$. However, in terms of the  parameter $g$, it appears as if Eq.~\eqref{eq:U} holds for arbitrarily large particle momentum exchange, $\vec{K} = \vec{p}-\vec{r}$. In other words,  the interaction in \eqref{eq:U} lacks of a characteristic momentum scale, which nevertheless exists for real  interactions but  depends on the microscopic details of the two-atom potential.

 The lack of a characteristic momentum scale  has important consequences for the calculation of physical properties using  the single-channel model.  For instance, the perturbation series for the ground state energy is plagued with divergences arising from the behavior of  integrals at high momenta (see e.g. ~\cite{abrikosov1975methods,pathria1996statistical}). 
 In addition,  for momenta smaller than the inverse effective range (i.e. $p\ll R^{-1}$~\footnote{This condition obeyed by most ultracold  gases even in the vicinity of  broad Feshbach resonances where $|a_s|$ diverges}), the momentum distribution, 
$n_{p\sigma} = \langle c^{\dag}_{\vec{p}\sigma} c_{\vec{p}\sigma}\rangle$ exhibits a $p^{-4}$ tail at large momenta $p\gg k_F$  ($k_F$ is the Fermi momentum). This  renders divergent the kinetic 
energy,   $E_{\text{kin}}=\sum_{p\sigma} n_{p\sigma} \epsilon_p$~\cite{1/k4tail,shinatan_contact} 
($\epsilon_p = p^2/2m$ is the single-particle dispersion and $m$ is the atom mass). In connection to this problem, Tan~\cite{shinatan_contact} has recently shown that the total energy can be entirely written in terms of the  momentum distribution $n_{k\sigma}$  and a parameter, $C$, which he  termed `contact'. The latter controls the asymptotic behavior $\sim p^{-4}$ of the momentum distribution $n_{p\sigma}$.

 In many-body perturbation theory, the  removal   (renormalization)  of the divergences described above  begins by recognizing that the coupling $g$ in Eq.~\eqref{eq:U} is not  physical and must be replaced by the  measurable $s$-wave scattering length, $a_s$. Thus, 
Eq.~\eqref{eq:U} can be used to compute the two-particle  scattering amplitude (see  Eq.~\ref{eq:tmatrix} below). We can parametrize the latter in terms of the $s$-wave phase shift   $\delta_s(K)$. In the limit of  small 
momentum exchange  of the colliding particles $\vec{K}=\vec{p}-\vec{r}$, the phase shift obeys:
\begin{equation}
K \cot\delta_s(K)  = -\frac{1}{a_s} + \frac{K^2R}{2}+\cdots 
\end{equation}
This expression allows to relate the coupling $g$ in Eq.~\eqref{eq:U} to the scattering length $a_s$ ($R$ is the effective range). An equivalent treatment relies on perturbation theory to relate $g$ to   $a_s$ through the following expression for the two-particle  $T$-matrix~\cite{abrikosov1975methods,pathria1996statistical}:
\begin{equation}
T_{\sigma\alpha}(\vec{k},\vec{p}) = g + \frac{2g^2}{\Omega}\sum_{\vec{qr}} \frac{\delta_{\vec{p}+\vec{k},\vec{q}+\vec{r}}}{E_{pkqr}} + O(g^3) = \frac{4\pi a_s}{m}.
\label{eq:tmatrix}
\end{equation}
In the above equation, $\sigma \neq \alpha$ and  $E_{pkqr} = \epsilon_{p}+\epsilon_{k}-\epsilon_{q}-\epsilon_{r}$.  

In this paper, we study how to remove the divergences that appear in the expressions describing the nonequilibrium dynamics of the total energy following an interaction quantum quench. We consider a fairly general  quench in which the system evolution is dictated by the Hamiltonian:
\begin{equation}
H(t) = H_0 + S(t) V,
\end{equation}
where $H_0 =\sum_{\vec{p}\sigma} \epsilon_p c^{\dag}_{\vec{p}\sigma} c_{\vec{p}\sigma}$ is the kinetic energy and 
$V$ is given in Eq.~\eqref{eq:U}. The function $S(t)$ 
describes the experimental protocol followed to turn on the interaction starting from the non-interacting system (accounting for the effects of  a weak initial interaction  can be also done perturbatively and will be reported elsewhere~\cite{unpub}). We  will see that the divergences that plague the calculation of the ground state energy in equilibrium are also present in the perturbative calculation of the total energy dynamics. 
Specializing  to the case of a linear ramp, where
\begin{equation}
S(t)=\theta(t)\left[\theta(T_r-t)\frac{t}{T_r}+\theta(t-T_r)\right],\label{eq:ramp_protocol}
\end{equation}
we obtain the time-evolution of the total energy and the momentum distribution as a function of ramp time $T_r$.
Tuning  $T_r$ allows us to study the crossover from the sudden (i.e. $T_r\to 0$) to the quasi-adiabatic (i.e. $T_r\to +\infty$) limit.

 In order to compute the total energy dynamics, we also compute the instantaneous momentum distribution, i.e.
\begin{align}
n_{p\sigma}(t)=\mathrm{Tr} \: \left[ \rho(t) c^{\dagger}_{\vec{p}\sigma}c_{\vec{p}\sigma} \right],
\end{align}
where $\rho(t)$ describes the state of the system at time $t$ (see Sec.~\ref{sec:appa} for details).
The perturbative expansion for this quantity does not 
contain any divergent integrals. However, the leading perturbative corrections appearing at $O(g^2)$ 
behave as $n_{p\sigma}\sim p^{-4}$ at large momentum $p$.  Likewise, the dynamics of the total energy
is obtained from:
\begin{align}
E_{\text{tot}}(t)&=\mathrm{Tr} \: \left[ \rho(t) H(t)\right] \notag \\
&= \mathrm{Tr} \: \left[ \rho(t) H_0\right]  + S(t) 
\mathrm{Tr} \: \left[ \rho(t) V\right],
\end{align}
where the first term in the second line is the instantaneous kinetic energy, i.e.
\begin{equation} 
E_{\mathrm{kin}}(t) =  \mathrm{Tr} \: \left[ \rho(t) H_0 \right]  = \sum_{\vec{p}\sigma} \epsilon_{p} n_{p\sigma}(t). 
\end{equation}
The high momentum tail  $\sim p^{-4}$ of the instantaneous momentum distribution $n_{p\sigma}(t)$ at $p\gg k_F$ renders $E_{\mathrm{kin}}(t)$ divergent, as in the equilibrium case. Interestingly, when
the coupling $g$ is replaced by the  scattering length $a_s$ by the renormalization method described in Sec.~\ref{sec:re}, the divergence  $E_{\mathrm{kin}}(t)$ is cancelled by an additional contribution from
the interaction energy $E_{\mathrm{int}}(t) = S(t) 
\mathrm{Tr} \: \left[ \rho(t) V\right]$. This 
cancellation parallels the similar cancellation happening in equilibrium and  allows us to
isolate the leading term in  perturbative expansion of the tail of the  instantaneous momentum distribution. For  a linear interaction ramp, we find that the definition of Tan's contact requires some care in order to fully capture the crossover behavior  of the tail  from the sudden to the adiabatic limit.

 Our results also allow to explore the  dynamics of the two-component Fermi gas following  
linear ramp in the interaction strength. A subject of particular interest in this situation is existence of a 
 pre-thermalized regime ~\cite{Moeckel2008,Moeckel2009,nessi_shorttime2014,
nopreth2d_2014,prethandth_2009_eckstein,Nessi_glass_2015,prethandth_silva_2013,turnable_integrablity_2014,nearintegrable_Lagen2016,Moeckel2010,spinprethermal2017,prethandth_mitra_2013,PhysRevLett.97.156403,preth_th_luttinger_2016,prethspinchain_Gong2013,GGEpreth_2011_kollar,Miguel2016,spin_prethandth_2015,nearintegrable_alba_2017,preth_spin_short_2018}. Thus, we have explored the existence of 
a pre-thermalized regime as a function of the ramp time. Previous studies on pre-thermalization in ultracold Fermi gases have focused on the behavior of the momentum distribution near the Fermi momentum $k_F$ at zero temperature~\cite{Moeckel2008,Moeckel2009,nessi_shorttime2014,nopreth2d_2014,prethandth_2009_eckstein,Nessi_glass_2015,GGEpreth_2011_kollar,Miguel2016,spin_prethandth_2015}. It was concluded that
the persistence  of a discontinuity at $k_F$  at the same  time that the total energy has reached its final (thermalized) value~\cite{PhysRevLett.93.142002} characterizes the pre-thermalized regime. 
Here we report results for the dynamics of  the full momentum distribution at finite temperatures as well as large momenta, which can provide  a more  experimentally accessible way to characterize the pre-thermalized regime. This is because  
in realistic systems, the discontinuity of the momentum distribution at $k_F$ is absent due to finite temperature effects and trap confinement. On the other hand, as we argue below, the dynamics of the full momentum distribution at finite temperature and its asymptotic behavior at high momenta  contains a great deal of useful information about the pre-thermalized regime. 
 
  The rest of this article is organized as follows. In section~\ref{sec:appa}, we give the derivation of observables. In section~\ref{sec:re}, we discuss the renormalization procedure in and out of equilibrium. In section~\ref{sec:ramp}, we specialize to the a linear ramp quench and study the dynamics of total energy and the momentum distribution to show the emergence of pre-thermalization in short time. In section~\ref{sec:contact}, we derive the contact for a ramp of interaction in the interaction  strength  and discuss its dynamics and relation to pre-thermalization. In Section~\ref{sec:conclu}, we summarize our results and present the conclusions of this work. Throughout, we use units where $\hbar=1$ and $k_B=1$.
\section{Evolution of Observables}\label{sec:appa}
\begin{figure}[t]
\center\includegraphics[width=0.9\columnwidth]{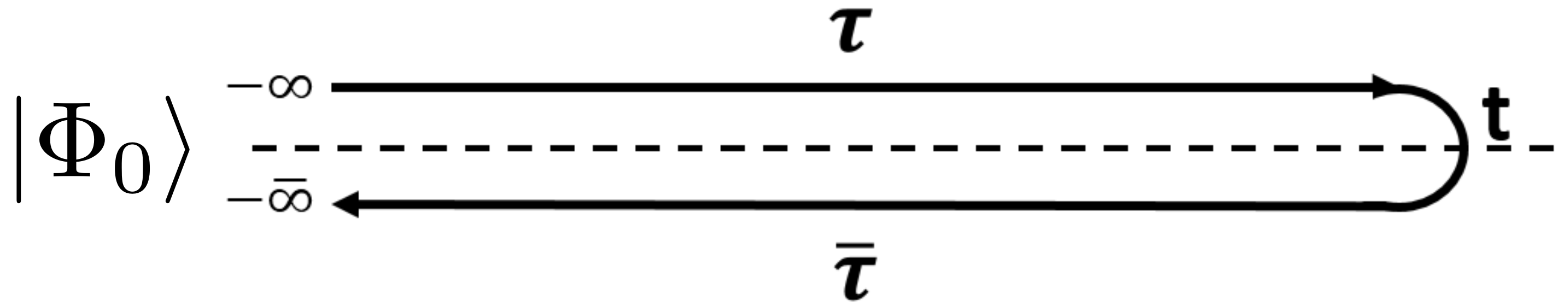}
\caption{\label{1} Closed-time contour $C$. Times $\tau$ and $\bar{\tau}$ lie on the time ordered and anti-time ordered branches, respectively. $\tau$ is earlier than $\bar{\tau}$ in contour ordering. The turning point $t$  is the time of at which observable in which we are interested is evaluated. $|\Phi_0\rangle$ is the initial state.}
\label{contour}
\end{figure}

 In what follows, we use perturbation theory to obtain the short to intermediate time dynamics of the system as the interaction is quenched. To leading order in the scattering length, $a_s$, this treatment is valid when the strength of interaction  being quenched is small. Indeed, earlier work~\cite{Moeckel2008,nessi_shorttime2014,Miguel2016,Moeckel2009,Moeckel2010,perturb_prethandth_2016,PhysRevB.92.235135,GGEpreth_2011_kollar,perturb_werner2013,kollarperturbation_2013}, it has been established  that pre-thermalization is accessible through perturbation theory in the quenched interaction. We calculated the observables to second order in the quench interaction. The valid time scale therefore fulfills the condition  $E_F t\sim (k_Fa_s)^{-2} \ll (k_Fa_s)^{-3}$, wheres $E_F = k^2_F/2m$ is the Fermi energy.  This requires $k_F a_s \ll 1$ in order to apply to hold at long times. 

  In this section, we provide the details of the derivation of the time evolution of the observables considered in this work. In the interaction picture,   the evolution of an observable can be calculated from the following expression, which is amenable to perturbative expansion:
\begin{align}
\langle O(t)\rangle&=\frac{\langle \mathcal{T}[e^{-i \int_C dt\: \tilde{V}(t)} O(t)]\rangle}{\langle \mathcal{T}[e^{-i \int_C dt\: \tilde{V}(t)}] \rangle}\label{eq:opert}.
\end{align}
In Eq.~\eqref{eq:opert}, $\tilde{V}(t)=e^{i H_0 t}Ve^{-i H_0 t}S(t)$ is the quench interaction in interaction picture, and $O = c^{\dag}_{\vec{p}\sigma}c_{\vec{p}\sigma},V,\ldots $ stands for the observable of interest.
Time $t$ lies on the closed contour $C$ shown in Fig.~\ref{contour} and $\mathcal{T}$ is the time-ordering symbol on $C$. Since we are working with a closed time contour,  trictly speaking, the denominator of Eq.~\eqref{eq:opert} equals unity. However, it is needed when expanding in powers of $\tilde{V}(t)$ in order to cancel disconnected Feynman graphs. The expectation values are computed according to
 $\langle ... \rangle = \text{Tr}[ e^{-\beta H_0} ...]$, which is the thermal average with respect to an non-interacting initial Hamiltonian at absolute temperature $T = \beta^{-1}$.

 Expanding  Eq.~\eqref{eq:opert} in powers of the quenched interaction $\tilde{V}(t)$ yields:
\begin{align}
O(t)&=\langle O\rangle\notag -i\int_C dt_1 \langle  \mathcal{T}\left[ \tilde{V}(t_1)O(t) \right] \rangle_c \notag \\
&\quad + \frac{(-i)^2}{2!}\int_C dt_1 dt_2 \, \langle  \mathcal{T} \left[ \tilde{V}(t_1)\tilde{V}(t_2)O(t) \right] \rangle_c\notag \\
&\quad + \cdots
\end{align}
As mentioned above, we shall only take into account the fully connected contributions (denoted by $\langle \ldots \rangle_c$ in the expression above) resulting from the application of Wick's theorem. When using the latter, there are four possible choices of time arguments for the fermion propagator:
\begin{equation}
i G_{p\sigma}(t_1,t_2)=  \langle \mathcal{T} \left[ c_{\vec{p}\sigma}(t_1)c_{\vec{p}\sigma}^{\dagger}(t_2)\right]\rangle, \label{eq:g0}
\end{equation}
where $t_1$ and $t_2$ can be either in the $\tau$ or $\bar{\tau}$ branches of $C$.  The free fermion propagator can be written in matrix form as follows:
\begin{align}
&\mathcal{G}_{p\sigma}(a,b)=
\begin{pmatrix}
iG^{T}_{p\sigma}(a,b)&iG^{<}_{p\sigma}(a,\bar{b})\\
iG^{>}_{p\sigma}(\bar{a},b)&iG^{\tilde{T}}_{p\sigma}(\bar{a},\bar{b})
\end{pmatrix}.\label{eq:g}
\end{align}
Using  $c_{\vec{p}\sigma}(t)=c_{\vec{p}\sigma}e^{-i\epsilon_{p}t}$, the 
elements of the above matrix can be evaluated to yield:
\begin{align}
i G_{p\sigma}^{<}(t_1,\bar{t}_2)&= -n_{p\sigma}^0e^{i\epsilon_p(t_2-t_1)},\label{eq:g1}\\
i G_{p\sigma}^{>}(\bar{t}_1,t_2)&=(1- n^0_{p\sigma})e^{i\epsilon_p(t_2-t_1)},\label{eq:g2}\\
i G^{T}_{p\sigma}(t_1,t_2)&=
\theta(t_1-t_2) iG_{p\sigma}^{>}(\bar{t}_1,t_2)\notag\\
&+\theta(t_2-t_1)iG^{<}_{p\sigma}(t_1,\bar{t}_2),\label{eq:g3}\\
i G^{\tilde{T}}_{p\sigma}(\bar{t}_1,\bar{t}_2)&=
\theta(t_2-t_1) iG^{>}_{p\sigma}(\bar{t}_1,t_2)\notag\\ 
&\quad +\theta(t_1-t_2)iG^{<}_{p\sigma}(t_1,\bar{t}_2).\label{eq:g4}
\end{align}

\subsection{Instantaneous momentum distribution}\label{sec:md}
\begin{figure}[b]
\center
\includegraphics[width=\columnwidth]{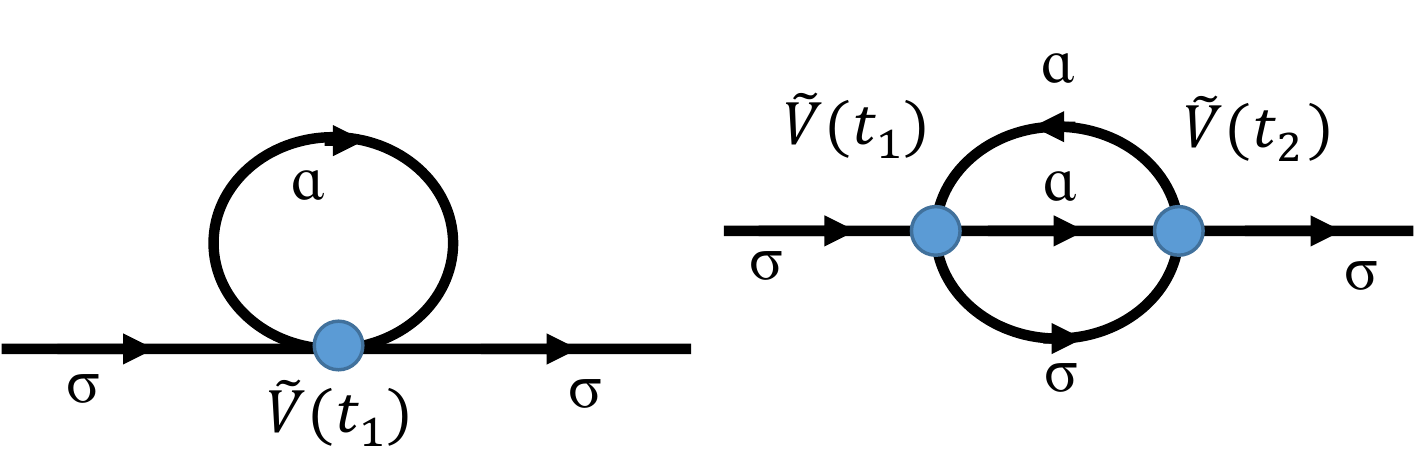}
\caption{First and second order diagram for momentum distribution. $\tilde{V}(t)$ is the quench interaction and its time dependence can be chosen to be on the two time branches. Therefore, we need to consider four types of green function from the contour ordering.}\label{fig:feyn1}
\end{figure}

 Let us first consider the evolution of  the momentum distribution. The latter is obtained from the Eq.~\eqref{eq:opert} by setting $O = \hat{n}_{p\sigma} = c^{\dag}_{p\sigma} c_{p\sigma}$. Hence,
expanding the evolution operator up to second order in the quenched interaction, the following expression is obtained:
\begin{multline}
n_{p\sigma}(t)=\langle \hat{n}_{p\sigma}\rangle\notag -i\int_C dt_1 \langle  \mathcal{T}\left[ \tilde{V}(t_1)\hat{n}_{p\sigma}(t) \right] \rangle_c  \\
+\frac{(-i)^2}{2!}\int_C dt_1 dt_2 \, \langle  \mathcal{T} \left[ \tilde{V}(t_1)\tilde{V}(t_2)\hat{n}_{p\sigma}(t) \right] \rangle_c 
+ \cdots 
\end{multline}
The expectation values in the above expressions can be represented diagrammatically. In the closed time contour, we can express the expectation value using  the self-energy and propagator matrices, which to second order in the quenched interaction yield:
\begin{multline}
n_{p\sigma}(t) = \langle \hat{n}_{p\sigma}\rangle + \int_C  dt_1\:   \mathcal{G}_{p\sigma}(t,t_1) \Sigma^{(1)}_{p\sigma}(t_1)\mathcal{G}_{p\sigma}(t_1,t) \\
+ \int_C
dt_1  \int_C dt_2 \: \mathcal{G}_{p\sigma}(t,t_2)\: 
 \Sigma^{(2)}_{p\sigma}(t_2,t_1) \mathcal{G}_{p\sigma}(t_1,t) 
 +\ldots \\
\label{eq:n}
\end{multline}
The propagator $\mathcal{G}_{p\sigma}(a,b)$ is defined in Eq.~\eqref{eq:g}, and $\Sigma^{(1)}_{p\sigma}(t_1)$, $\Sigma^{(2)}_{p\sigma}(t_2,t_1)$ can be found using diagrams (see below). For the calculation of equal-time expectation values, we choose the time argument   of the  observable (i.e. $t$) to lie slightly before the turning point of the contour $C$,  which is on the time ordered (i.e. $\tau$) branch. In this 
case, the fermion propagators must be obtained from 
Eq.~\eqref{eq:g0}, which yields:
\begin{equation}
\mathcal{G}_{p\sigma}(t,b)%
=e^{-i\epsilon_{p}(t-b)}\begin{pmatrix}
 1-n^0_{p\sigma}&-n^0_{p\sigma}\\0&0\end{pmatrix}
\label{eq:Gg1},
\end{equation}
where the non-vanishing entries correspond to either $b$ lying before or after $t$ on the contour $C$. Similarly,
\begin{equation}
\mathcal{G}_{p\sigma}(a,t)=
e^{-i\epsilon_{p}(a-t)}\begin{pmatrix} -n^0_{p\sigma}&0\\1-n^0_{p\sigma}&0\end{pmatrix}.
\label{eq:G2}
\end{equation}
The two non-zero entries in the above matrix correspond to $a$ lying before or after $t$ on the contour $C$. 

The self-energy can be calculated from the diagrams shown in Fig.~\ref{fig:feyn1} and the propagators, Eqs.~\eqref{eq:g1} to \eqref{eq:g4}.  Thus, to first order in $V(t)$, we obtain:
\begin{align}
&\Sigma^{(1)}_{\sigma}=\frac{g}{2\Omega}
\sum_{\vec{k}}n^0_{k,-\sigma}.\label{eq:s1}
\end{align}
Combining the matrix from Eq.~\eqref{eq:n},  the propagators, Eq.~\eqref{eq:Gg1} and Eq.~\eqref{eq:G2}, and the self-energy for the first order correction, Eq.~\eqref{eq:s1}, we obtain that first order correction to the instantaneous momentum distribution vanishes, i.e. $n_{p\sigma}^{(1)}(t)=0$. Ultimately, this is a consequence of the initial state being an eigenstate of the occupation operator $\hat{n}_{p\sigma} = c^{\dag}_{\vec{p}\sigma} c_{\vec{p}\sigma}$.

 At second order in the interaction, we need to use the following self-energy matrix, which contains four different combinations of the time arguments $(t_1,t_2)$ on the two branches of the  contour $C$:
\begin{align}
&\Sigma^{(2)}_{p\sigma}(b,a)=-\frac{2 g^2}{\Omega^2}S(t_1)S(t_2) \sum_{\vec{kqr}}\delta_{\vec{p}+\vec{k},\vec{q}+\vec{r}}\notag \\
&\times
\begin{pmatrix}\bar{\Sigma}_{\sigma}^{(2,T)}(b,a)&\bar{\Sigma}_{\sigma}^{(2,>)}(\bar{b},a)\\\bar{\Sigma}_{\sigma}^{(2,<)}(b,\bar{a}) &\bar{\Sigma}_{\sigma}^{(2,\tilde{T})}(\bar{b},\bar{a})\end{pmatrix}.
\end{align}
Evaluating the diagrams in Fig.~\ref{fig:feyn1} using the  propagators in Eqs.~\eqref{eq:g1} to \eqref{eq:g4}, we obtain the following expressions for the elements of the self-energy matrix:
\begin{align}
 \bar{\Sigma}_{\sigma}^{(2,<)}(t_2,\bar{t}_1)&=i^3G^{<}_{k\alpha}(t_2,\bar{t}_1)G^{>}_{q\sigma}(\bar{t}_1,t_2)G^{>}_{r\alpha}(\bar{t}_1,t_2),\notag\\
&=-(1-n^0_{q\sigma})(1-n^0_{r\alpha}) n^0_{k\alpha}e^{i(t_1-t_2)(\epsilon_{q}+\epsilon_{r}-\epsilon_{k})} ,\label{s1}\\
\bar{\Sigma}^{(2,>)}(\bar{t}_2,t_1)&=i^3G^{>}_{k\alpha}(\bar{t}_2,t_1)G^{<}_{q\sigma}(t_1,\bar{t}_2)G^{<}_{r\alpha}(t_1,\bar{t}_2),\notag \\
  &=n^0_{q\sigma}n^0_{r\alpha}(1-n^0_{k\alpha})e^{i(t_1-t_2)(\epsilon_{q}+\epsilon_{r}-\epsilon_{k})},\label{s2}\\        
 \bar{\Sigma}^{(2,T)}(t_2,t_1)&=i^3G^{T}_{k\alpha}(t_2,t_1)G^{T}_{q\sigma}(t_1,t_2)G^T_{r\alpha}(t_1,t_2),\notag\\
 =\theta(t_2&-t_1)\Sigma^{(2,>)}(t_2,\bar{t}_1)+\theta(t_1-t_2)\Sigma^{(2,<)}(\bar{t}_2,t_1),\label{s3} \\ 
 \bar{\Sigma}^{(2,\tilde{T})}(\bar{t}_2,\bar{t}_1)&=i^3G^{\bar{T}}_{k\alpha}(\bar{t}_2,\bar{t}_1)G^{\bar{T}}_{q\sigma}(\bar{t}_1,\bar{t}_2)G^{\bar{T}}_{r\alpha}(\bar{t}_1,\bar{t}_2),\notag\\
=\theta(t_1&-t_2)\Sigma^{(2,>)}(t_2,\bar{t}_1)+\theta(t_2-t_1)\Sigma^{(2,<)}(\bar{t}_2,t_1).\label{s4}
\end{align}
Combining Eq.~\eqref{eq:n}, the propagators (cf. Eq. ~\ref{eq:g1} to Eq.~\ref{eq:g4}) and the second order corrections to the self-energy, Eqs.~\eqref{s1} to \eqref{s4},  we arrive at:
\begin{equation}
n^{(2)}_{p\sigma}(g,t) =-\frac{2g^2}{\Omega^2}\sum_{\vec{kqr}} A^{\sigma}_{pkqr} F^{(2)}(E_{pkqr},t) \delta_{\vec{p}+\vec{k},\vec{q}+\vec{r}},\label{eq:nk2}
\end{equation}
where $A^{\sigma}_{pkqr}$ denotes the function:
\begin{multline}
A^{\sigma}_{pkqr} = \sum_{\alpha\neq\sigma}\left[n^0_{p\sigma}n^0_{k\alpha}(1-n^0_{q\sigma})(1-n^0_{r\alpha}) \right. \\ 
\left. \quad   -(1-n^0_{p\sigma})(1-n^0_{k\alpha})n_{q\sigma}^0n_{r\alpha}^0\right].
\end{multline}
In Eq.~\eqref{eq:nk2},  $E_{pkqr}=\epsilon_p+\epsilon_k-\epsilon_q-\epsilon_r$, and the function 
\begin{equation}
F^{(2)}(E,t) = \int\limits_{-\infty}^{t} dt_1 \int\limits_{-\infty}^{t} dt_2\:    S(t_1)S(t_2) \: e^{i E(t_1-t_2)} \label{eq:f2}
\end{equation}
has been introduced.  Note that $F^{(2)}(E,t)$ depends on the explicit form of $S(t)$ (see Sec.~\ref{sec:ramp} and Appendix~\ref{sec:F} for the form of this function in a number of important limiting cases).  

For $p > k_F$, the above expression, Eq.~\eqref{eq:nk2}, reduces to
\begin{multline}
n^{(2)}_{p\sigma}(g,t) =\frac{2g^2}{\Omega^2}\sum_{\vec{kqr},\sigma} (1 -n_{k,-\sigma}) n_{q,\sigma}n_{r,-\sigma}\\
\times F^{(2)}(E_{pkqr},t) \delta_{\vec{p}+\vec{k},\vec{q}+\vec{r}}.
\end{multline}
Furthermore, momentum conservation requires that $\vec{k} = \vec{q}+\vec{r} - \vec{p}$. Since the occupation factors
in this expression  force $q, r \leq k_F$, it follows that $|\vec{q}+\vec{r}| \leq 2 k_F$. Thus,  for $p\gg k_F$, $k \gg k_F$, and
the above expression simplifies to
\begin{multline}
n^{(2)}_{p\gg k_F,\sigma}(g, t)    =\frac{2g^2}{\Omega^2}\sum_{\vec{kqr}}  n_{q\sigma}n_{r,-\sigma}  \delta_{\vec{p}+\vec{k},\vec{q}+\vec{r}}\\
\times F^{(2)}(E_{pkqr},t).\label{eq:nplarge}
\end{multline}
It will be shown in Sec.~\ref{sec:contact} that this expression leads to a $\sim p^{-4}$ tail of the momentum distribution, which
renders the kinetic energy divergent. 
\subsection{Unrenormalized kinetic energy}\label{sec:Ekin}

 From the above result for the instantaneous momentum distribution, we can derive the second order 
 correction to the kinetic energy:
\begin{multline}
\delta E_{\mathrm{kin}}^{(2)}(q,t)=\sum_{p,\sigma}\epsilon_p n^{(2)}_{p\sigma}(t)\\
=-\frac{2g^2}{\Omega^2}\sum_{\vec{pkqr},\sigma} \epsilon_p A^{\sigma}_{pkqr} F^{(2)}(E_{pkqr},t)\delta_{\vec{k}+\vec{p},\vec{q}+\vec{r}} \\
=\frac{-g^2}{2\Omega^2}\sum_{\vec{pkqr},\sigma}  A^{\sigma}_{pkqr} E_{pkqr}F^{(2)}(E_{pkqr},t)\delta_{\vec{k}+\vec{p},\vec{q}+\vec{r}}
\label{eq:Ek}
\end{multline}
where, in the last line,  we have used that $\sum_{\sigma}A^{\sigma}_{pkqr} = \sum_{\sigma}A^{\sigma}_{kpqr}  = - \sum_{\sigma}A^{\sigma}_{qrpk} =  -\sum_{\sigma}A^{\sigma}_{rqpk} $ and $F^{(2)}(-E,t) = F^{(2)}(E,t)$. Notice that in  two extreme limits, i.e. the sudden and adiabatic limits  $F^{(2)}(E,t) \sim 1/E$ for $E\to +\infty$ (see Appendix~\ref{sec:F}). Thus,  the above expression  is divergent, as in equilibrium (the expression of the adiabatic limit coincides with the equilibrium one). Generically, for any other quench protocol function $S(t)$ between these two extreme limits,  we expect the same behavior and the above expression to remain divergent, as it is also confirmed for a linear ramp quench, see Sec.~\ref{sec:ramp}.
\subsection{Unrenormalized interaction energy}\label{sec:Eint}
\begin{figure}[b]
\includegraphics[width=\columnwidth]{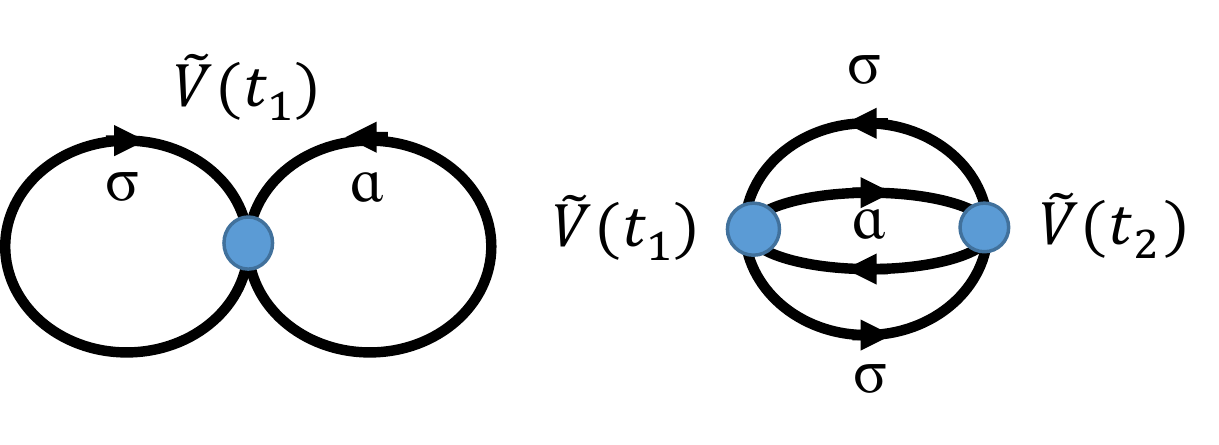}
\caption{Feynman diagrams for the first- and second-order corrections to the interaction energy.}
\label{fig:intdiag}
\end{figure}
 The interaction energy can be obtained from Eq.~\eqref{eq:opert} 
by setting $O = \tilde{V}(t)$. Expanding this expression to second order in $g$ yields
\begin{multline}
\delta E_{\mathrm{int}}(g,t)=\langle \Phi_0|\tilde{V}(t)| \Phi_0\rangle\\
-i\int_C   dt_1 \: \langle \Phi_0| \mathcal{T}\left[\tilde{V}(t_1)\tilde{V}(t)\right]|\Phi_0\rangle_c + \cdots 
\end{multline}
The calculation in this case proceeds in a  similar fashion to the one in previous section. Evaluating the Feynman  diagrams in Fig.~\ref{fig:intdiag}, we obtain:
\begin{align}
\delta E_{\mathrm{int}}(t)&=V^{(1)}(t) + V^{(2)}(t) + \cdots,\\
V^{(1)}(t) &= \frac{g}{2\Omega}S(t)\sum_{\alpha\neq\sigma}
\sum_{kp}n^0_{k\sigma}n^0_{p\alpha}\label{E1},\\ 
V^{(2)}(t) &= \int_Cd\tau_1\: S(\tau_1) 
\Lambda^{(2)}(\tau_1;t). 
 \label{eq:int}
\end{align}
Note that the first order  term,  $V^{(1)}(t)$,  Eq.~\eqref{E1}, is proportional to the  function $S(t)$. 
The second order term, $V^{(2)}(t)$ can be obtained from the matrix:
\begin{align}
\Lambda^{(2)}(t_1,t)&=\begin{pmatrix}\Lambda^{(2,T)}(t_1,t)&\Lambda^{(2,<)}(t_1,\bar{t})\\
\Lambda^{(2,>)}(\bar{t}_1,t)&\Lambda^{(2,\tilde{T})}(\bar{t}_1,\bar{t})\end{pmatrix},\\
&=\begin{pmatrix}\Lambda^{(2,T)}(t_1,t)&0\\
\Lambda^{(2,>)}(\bar{t}_1,t)&0\end{pmatrix}.
\end{align}
In the last line, we have used  that $t$ lies on the time-ordered branch of the contour $C$. Using the free propagator (cf. Eqs.~\ref{eq:g1} to \ref{eq:g4}), the second order term becomes:
\begin{equation}
\begin{aligned}
&\Lambda^{(2)}(t_1,t)=\frac{-ig^2 S(t)}{\Omega^2}\sum_{\vec{pkqr}}e^{iE_{pkqr}(t_1-t)}\\
&\times \begin{pmatrix}
(1-n^0_{p\sigma})(1-n^0_{k\alpha})n^0_{q\sigma}n^0_{r\alpha}&0 \\
n^0_{p\sigma}n^0_{k\alpha}(1-n^0_{q\sigma})(1-n^0_{r\alpha})&0\end{pmatrix}.\label{eq:L2}
\end{aligned}
\end{equation}
From the above expression, using Eq.~\eqref{eq:int}, we obtain the second order correction to the interaction energy:
\begin{multline}
\delta E_{\mathrm{int}}^{(2)}(g,t)=S(t)\int_C dt_1\: \Lambda^{(2)}(t,t_1) \: S(t_1) \\
=\frac{g^2 S(t)}{\Omega^2}\sum_{\vec{pkqr},\sigma} A^{\sigma}_{pkqr}F^{(1)}(E_{pkqr},t)\delta_{\vec{k}+\vec{p},\vec{q}+\vec{r}}.\label{eq:eint}
\end{multline}
In the last line, we  have introduced the function
\begin{equation}
F^{(1)}(E,t)=\int_{-\infty}^{t} \sin\left[E(t-t_1)\right] S(t) dt_1, \label{eq:f1}
\end{equation}
which  depends on the function $S(t)$ that defines the quench protocol. The sine function in Eq.~\eqref{eq:f1} appears after  swapping the dummy momenta around  in order to symmetrize the integrand, that is,  after using:
\begin{align}
\sum_{\vec{pkqr}}f_{\vec{pkqr}}=\frac{1}{2}\sum_{\vec{pkqr}}\left[ f_{\vec{pkqr}}+f_{\vec{qrpk}} \right].
\end{align}
Again, notice that that in  the sudden and adiabatic limits,  $F^{(1)}(E,t) \sim 1/E$ for $E\to +\infty$ (see Appendix~\ref{sec:F}), which means that  Eq.~\eqref{eq:eint} is divergent, as in equilibrium (the expression of the adiabatic limit coincides with the equilibrium result, see Appendix.~\ref{sec:F}). Generically, for any other quench protocol function $S(t)$ between these two extreme limits,  we expect the above expression to remain divergent. This  is explicitly confirmed for a linear ramp protocol in Sec.~\ref{sec:ramp}.

\subsection{Unrenormalized total energy}\label{sec:Etot}
From the results obtained in  previous sections for the interaction and kinetic energy, we  can obtain the dynamics of the total energy  for a  quench in which the interaction is switched on according to an arbitrary protocol described by $S(t)$. The first order correction is:
\begin{equation}
\delta E^{(1)}_{\mathrm{tot}}(g,t)=\frac{g}{2\Omega}\sum_{\vec{pk},\sigma\neq\alpha} n_{p\sigma}n_{k\alpha},
\end{equation}
and the second order correction reads:
\begin{align}
\delta E^{(2)}_{\mathrm{tot}}(g,t)&=\delta E^{(2)}_{\mathrm{kin}}(g,t)+\delta E^{(2)}_{\mathrm{int}}(g,t),\notag\\
&=\frac{g^2}{\Omega^2}\sum_{\vec{pkqr},\sigma} A^{\sigma}_{pkqr} \: F_{\mathrm{tot}}(E_{pkqr},t) \delta_{\vec{p}+\vec{k},\vec{q}+\vec{r}}, \notag\\
&= \frac{4g^2}{\Omega^2}\sum_{pkqr}n^0_{p,\uparrow}n^0_{k,\downarrow}(1-n^0_{q,\uparrow})(1-n^0_{r,\downarrow})\notag\\&\qquad\times F_{\mathrm{tot}}(E_{pkqr},t) \delta_{\vec{p}+\vec{k},\vec{q}+\vec{r}}
\label{eq:Etott}
\end{align}
where, in the last line, we have introduced the function:
\begin{align}
F_{\mathrm{tot}}(E,t)=S(t)F^{(1)}(E,t)-\frac{E}{2}F^{(2)}(E,t).\label{eq:F}
\end{align}
and used that $F_{\mathrm{tot}}(-E,t) = -F_{\mathrm{tot}}(E,t)$ to obtain the expression in the last line of \eqref{eq:Etott}.
Furthermore, notice that, as shown in Appendix~\ref{sec:F},  $F_{\mathrm{tot}}(E,t)$  vanishes in the limit of a sudden quench. This ensures that the second order correction to the total energy vanishes, as required by energy conservation. See Eq.~\eqref{eq:zero} in Appendix~\ref{sec:F} and Sec.~\ref{sec:ramp} for a more in depth discussion of this point.   

\section{Elimination of  divergences}\label{sec:re}

As we have mentioned in the Introduction, perturbation theory in the powers of the \emph{bare} interaction $V\propto g$ (cf. Eq.~\ref{eq:U}) yields an expression for the equilibrium total
energy  containing divergent integrals~\cite{Abrikosov1965,Parthia2011}. The divergences appear at second in the coupling $g$ and their elimination, i.e. their `renormalization', is possible by realizing that the \emph{unphysical} $g$ must be replaced by the \emph{physical} $s$-wave scattering length, $a_s$. The latter  is related to $g$ via the two-particle scattering amplitude (i.e. the $T$-matrix), see Eq.~\eqref{eq:tmatrix}. 

The same divergences reappear in the perturbative
expansion for the dynamics of the total energy obtained in  previous sections. Indeed, as we show by explicit calculation
below, the  kinetic energy is divergent because  the leading order [i.e. $O(g^2)$] correction to instantaneous momentum distribution  behaves as $\sim p^{-4}$ for $p\gg k_F$. Hence,   $E_{\mathrm{kin}}(t) = \sum_{p\sigma} \epsilon_p n_{p\sigma}(t)$ is divergent at all times $t$. Similar divergences appear in the  expressions for the interaction energy at $O(g^2)$ as well.  

  However, in the nonequilibrium case,  it is difficult to rely on the   $T$-matrix because defining the latter requires  the introduction of asymptotic scattering states. Those states are well defined for interactions that are switched on and off adiabatically, as it is assumed in  equilibrium, but  this becomes difficult when the interaction changes (sometimes, very rapidly) in time as it is the case of our study. Thus,  the generalization of the renormalization procedure employed in equilibrium to the problem of interest here is not straightforward. Instead,  we show below that the renormalization procedure can be carried out by computing the perturbative corrections to the evolution of the total two-particle energy.

\subsection{Renormalization in equilibrium}
 Let us begin by reviewing how the divergences are eliminated in  equilibrium case. As mentioned above, we shall compute the  shift of the total energy for two particles to relate $g$ to the $s$-wave scattering length, $a_s$~\cite{pathria1996statistical}.  This approach  reproduces the well-known equilibrium results based on the scattering matrix approach~\cite{pathria1996statistical,Abrikosov1965}. The shift to the total energy for two-particles is defined by:
\begin{align}\label{eq:eq}
\delta E^{2-\mathrm{body}}_{\mathrm{tot}}
(\vec{p}\sigma,\vec{k}\alpha)&=  E^{2-\mathrm{body}}_{\mathrm{kin}} + E^{2-\mathrm{body}}_{\mathrm{int}}\notag\\
&=\frac{\langle \Psi_{\vec{kp},\sigma\alpha} | \left(  H_0+V \right) | \Psi_{\vec{kp},\sigma\alpha}\rangle}{\langle \Psi_{\vec{kp},\sigma\alpha}  | \Psi_{\vec{kp},\sigma\alpha}\rangle}\notag \\
&\quad -\langle\Psi^{0}_{\vec{kp},\sigma\alpha}|  H_0 | \Psi^{0}_{\vec{kp},\sigma\alpha}\rangle,
\end{align}
where $|\Psi^{0}_{\vec{kp},\sigma\alpha}\rangle=c^{\dag}_{\vec{p}\sigma}c^{\dag}_{\vec{k}\alpha}|0\rangle$  describes a state of two free particles with $\sigma \neq \alpha$, and $|\Psi_{\vec{pk},\sigma\alpha}\rangle$ is the  two-particle state perturbed by the interactions. Using  time-independent perturbation theory, 
\begin{align}
|\Psi_{\vec{kp},\sigma\alpha}\rangle &=|\Psi^0_{\vec{kp},\sigma\alpha}\rangle +\sum_{\vec{qr}\neq \vec{kp}} |\Psi^{0}_{\vec{qr},\sigma\alpha}\rangle \notag\\ 
& \times \frac{\langle\Psi^{0}_{\vec{qr},\sigma\alpha} |V|\Psi^0_{\vec{kp},\sigma\alpha}\rangle}{E_{pkqr}}+O(V^2),\notag\\
&=|\Psi^0_{\vec{kp},\sigma\alpha}\rangle +\frac{g}{\Omega}\sum_{\vec{qr}} \frac{\delta_{\vec{p}+\vec{k},\vec{q}+\vec{r}} }{E_{pkqr}}  \left[ |\Psi^{0}_{\vec{qr},\sigma\alpha}\rangle \right. \notag\\
&\left.\qquad + |\Psi^{0}_{\vec{qr},\alpha \sigma}\rangle  \right] +  O(g^2),
\label{eq:2bodystate}
\end{align}
where  $E_{kpqr} = \epsilon_k + \epsilon_p - \epsilon_q - \epsilon_r$. Hence,  the shift of the kinetic energy is 
\begin{align}
 E^{2-\mathrm{body}}_{\mathrm{kin}} &= 
\frac{\langle \Psi_{\vec{kp},\sigma\alpha} |  H_0 | \Psi_{\vec{kp},\sigma\alpha}\rangle}{\langle \Psi_{\vec{kp},\sigma\alpha}  | \Psi_{\vec{kp},\sigma\alpha}\rangle}-\langle\Psi^{0}_{\vec{kp},\sigma\alpha}|  H_0 | \Psi^{0}_{\vec{kp},\sigma\alpha}\rangle \notag\\
&= -\frac{2g^2}{\Omega^2}\sum_{\vec{qr}\neq\vec{kp}} \frac{\delta_{\vec{k}+\vec{p},\vec{q}+\vec{r}}}{E_{pkqr}} + O(g^3).
\end{align}
To the same order in $g$, the shift to the interaction
energy reads:
\begin{align}
 E^{2-\mathrm{body}}_{\mathrm{int}} &= 
\frac{\langle \Psi_{\vec{kp},\sigma\alpha} |  V | \Psi_{\vec{kp},\sigma\alpha}\rangle}{\langle \Psi_{\vec{kp},\sigma\alpha}  | \Psi_{\vec{kp},\sigma\alpha}\rangle}\notag \\
&= \frac{g}{\Omega} +  \frac{4g^2}{\Omega^2} \sum_{\vec{qr}\neq \vec{kp}} \frac{\delta_{\vec{k}+\vec{p},\vec{q}+\vec{r}}}{E_{pkqr}} + O(g^3).\label{eq:perint}
\end{align}
Thus, the total energy shift is
\begin{align}
\delta E^{2-\mathrm{body}}_{\mathrm{tot}} &= E^{2-\mathrm{body}}_{\mathrm{kin}} + E^{2-\mathrm{body}}_{\mathrm{int}} \notag \\
&= \frac{g}{\Omega} +  \frac{2g^2}{\Omega^2} \sum_{\vec{qr}} \frac{\delta_{\vec{k}+\vec{p},\vec{q}+\vec{r}}}{E_{pkqr}} + O(g^3).
\label{eq:pertres}
\end{align}
Let us define the physical scattering amplitude by requiring that:
\begin{align}
\delta E^{2\mathrm{-body}}_{\mathrm{tot}}(p\sigma,k\alpha)=\frac{4\pi a_s}{m \Omega},\label{eq:as}
\end{align}
after equating it to Eq.~\eqref{eq:pertres}, we arrive at Eq.~\eqref{eq:tmatrix}. Inverting the series gives the  coupling $g$ in terms of the scattering length
\begin{align}
g=\frac{4\pi a_s}{m}-\frac{2}{\Omega}\left(\frac{4\pi a_s}{m}\right)^2 \sum_{qr}\frac{\delta_{\vec{p}+\vec{k},\vec{q}+\vec{r}}}{E_{pkqr}}+O(a_s^3),
\label{eq:invertg}
\end{align}
which allows to remove the divergences 
in the expressions for the many-particle ground
state energy~\cite{abrikosov1975methods,pathria1996statistical}. Note that the same result can be  obtained directly from the perturbative expression for the \emph{total} energy shift~\cite{,pathria1996statistical}.  However, we have chosen this more cumbersome method in order to emphasize that using the interaction energy instead would introduce in Eq.~\eqref{eq:invertg} a spurious factor of two (compare Eqs.~\eqref{eq:perint} and \eqref{eq:pertres}), which would not remove the divergences.  

\subsection{Nonequilibrium renormalization}\label{sec:reneq}
 
Let us now turn to the nonequilibrium situation. In the calculation described in Sec.~\ref{sec:appa}, the evolution of the total energy shift  is obtained in a perturbative series in the coupling $g$:
\begin{align}
\delta E_{\mathrm{tot}}(g,t)=E_{\mathrm{tot}}^{(1)}(g,t)+E_{\mathrm{tot}}^{(2)}(g,t)+ \cdots
\end{align}
where $E^{(n)}_{\mathrm{tot}}(g,t) = O(g^n)$. However, 
this expansion contains divergent integrals. 
Our goal in this section is to remove the divergences, which can be carried out after replacing $g$  by $a_s$, 
in a procedure similar to the one outlined above for the equilibrium case.

 Adding up the first and second order corrections  obtained in Sec.~\ref{sec:appa} yields:
\begin{multline}
\delta  E_{\mathrm{tot}}(g,t)=\frac{g }{\Omega}\sum_{\vec{pk}}n_{p,\uparrow}^0n_{k,\downarrow}^0 S(t)\\
+\frac{4g^2}{\Omega^2}\sum_{\vec{pkqr}}n^0_{p,\uparrow}n^0_{k,\downarrow}
(1-n^0_{q,\uparrow}) (1-n^0_{r,\downarrow}) \delta_{\vec{p}+\vec{k},\vec{q}+\vec{r}}\\
\times F_{\mathrm{tot}}(E_{pkqr},t)  +O(g^3).\label{eq:Etot}
\end{multline}
For two-particles, the evolution of the total energy shift can be obtained in a similar fashion, which leads to:
\begin{multline}
\delta E^{\mathrm{2-body}}_{\mathrm{tot}}(\vec{p}\sigma,\vec{k}\alpha,t)=\frac{gS(t)}{\Omega}+\frac{4g^2}{\Omega^2}\sum_{\vec{qr}}
\delta_{\vec{p}+\vec{k},\vec{q}+\vec{r}}\\
\times F_{\mathrm{tot}}(E_{\mathrm{pkqr}},t) +O(g^3).
\end{multline}
Next, we require that:
\begin{equation}
\delta E^{\mathrm{2-body}}_{\mathrm{tot}}(\vec{p}\sigma,\vec{k}\alpha,t) =  
\frac{4\pi a_s }{m} \frac{S(t)}{\Omega}.
\end{equation}
Inverting the series will gives the  coupling $g$ in terms of the scattering length~\footnote{Strictly speaking the coupling $g$ should be (weakly) time dependent. This is reminiscent of the equilibrium situation where Eq.~\ref{eq:invertg} would require $g$ to be (weakly) energy dependent.}:
\begin{align}
g(t)S(t)&=\left(\frac{4\pi a_s}{m}\right) S(t) -\frac{4}{\Omega}\sum_{\vec{qr}}\left(\frac{4\pi a_s}{m}\right)^2 \delta_{\vec{p}+\vec{k},\vec{q}+\vec{r}} \notag \\
& \times F_{\mathrm{tot}}(E_{pkqr},t) +O(a_s^3),
\label{eq:gc}
\end{align}
In the limit where the interaction is switched  adiabatically, notice that
$F_{\text{tot}}(E)=2E^{-1}$ (see Appendix~\ref{sec:F}) and  thus we recovers the equilibrium result, Eq.~\eqref{eq:invertg}.

\begin{figure*}
\centering
\includegraphics[width=\textwidth]{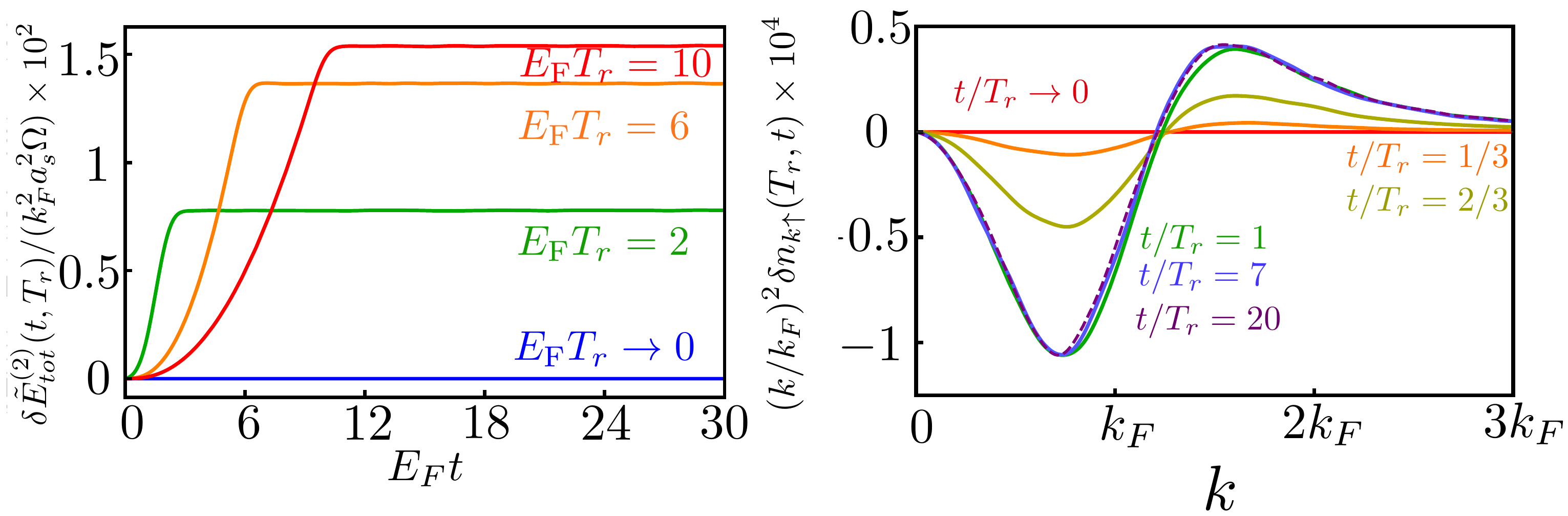}
\caption{(color line) (Left panel)  Dynamics of the second order correction to the total energy density (after renormalization), $\delta \tilde{E}^{(2)}_{\mathrm{tot}}(t,T_r)/\Omega$  for  different interaction ramp times $T_r$. Note that the second order correction vanishes as the sudden-quench limit (i.e. $T_r \to 0$) is approached, see discussion in Sec.~\ref{sec:ramp}. (Right panel) Dynamics of the momentum distribution after a linear ramp of the interaction strength.  We observe the development of a stationary distribution for $k\sim k_F$, which is characteristic of  a pre-thermalized regime. The dashed line corresponds to the largest time of the series, i.e. $t/T_r = 20$, and we have chosen $E_F T_r = 6$. The interaction strength is  $ a_s=0.02 k^{-1}_{\mathrm{F}}$ ($k_F$ is the Fermi momentum). The initial gas temperature  $T=0.1 T_F$, where $T_F$ is the Fermi temperature. } \label{fig:ramp}
\end{figure*} 

 Substituting Eq.~\eqref{eq:gc} into the first order correction for total energy shift, Eq.~\eqref{eq:Etot}, yields the following correction:
 \begin{align}
 \Delta_{\mathrm{tot}}(t)=\frac{4}{\Omega}\left(\frac{4\pi a_s}{m}\right)^2\sum_{\vec{pkqr}}n_{p,\uparrow}^0n_{k,\downarrow}^0 \delta_{\vec{p}+\vec{k},\vec{q}+\vec{r}}\notag\\
\times F_{\mathrm{tot}}(E_{pkqr},t),
 \end{align}
which is second order in the scattering length $a_s$ and  divergent. However, this divergence exactly cancels the one already present in the second order correction to the energy, i.e.  $E_{\mathrm{tot}}^{(2)}(g^2,t)$ (after replacing $g$  by $(4\pi a_s/m)$). Thus,  the finite expression for the dynamics of the total energy to second order in $a_s$ is :
\begin{multline}
\delta E_{\mathrm{tot}}\left(g = \frac{4\pi a_s}{m},t\right)-\Delta_{\mathrm{tot}}(t)=\left(\frac{4\pi a_s}{m}\right)\frac{ S(t)}{\Omega}\sum_{\vec{pk}}n_{p,\uparrow}^0n_{k,\downarrow}^0\\
-2\left(\frac{4\pi a_s}{m}\right)^2\sum_{\vec{pkqr}}n^0_{p,\uparrow}n^0_{k,\downarrow}\left(n^0_{q,\uparrow}+n^0_{r,\downarrow}\right) \delta_{\vec{p}+\vec{k},\vec{q}+\vec{r}}\\
\times F_{\mathrm{tot}}(E_{pkqr},t) +O(a_s^3),\label{eq:Etot1}
\end{multline}
 Furthermore,  according to the dependence on $S(t)$ and time,   we can  split 
$\Delta_{\mathrm{tot}}(t)$ into  the sum of $\Delta_{\mathrm{kin}}(t) \sim F^{(2)}(E,t)$ and $\Delta_{\mathrm{int}}(t)\sim F^{(1)}(E,t)$. Explicitly, 
\begin{align}
\Delta_{\mathrm{kin}}(t)&= -\frac{4}{\Omega^2}\left(\frac{4\pi a_s}{m}\right)^2\sum_{\vec{pkqr}}n^0_{p\uparrow}n^0_{k\downarrow}\delta_{\vec{p}+\vec{k},\vec{q}+\vec{r}}\notag\\
&\times \frac{E_{pkqr}}{2}F^{(2)}(E_{pkqr},t),\label{eq:dk}\\
\Delta_{\mathrm{int}}(t) &=\frac{4}{\Omega^2}\left(\frac{4\pi a_s}{m}\right)^2\sum_{\vec{pkqr}}n^0_{p\uparrow}n^0_{k\downarrow}\delta_{\vec{p}+\vec{k},\vec{q}+\vec{r}}\notag\\
&\times S(t) F^{(1)}(E_{pkqr},t).\label{eq:di}
\end{align}
Symmetrizing the dependence on momenta $\vec{p},\vec{k},\vec{q}$ and $\vec{r}$  of  the integrand with the 
help of $F^{(2)}(-E,t)  = F^{(2)}(E,t)$, $\Delta_{\text{kin}}(t)$ can be written as 
\begin{equation}
\Delta_{\text{kin}}(t) = \sum_{\vec{k}} \epsilon_k \delta n_{k}(t),\label{eq:renkin}
\end{equation}
where 
\begin{align}
\delta n_k (t) &=-\frac{4}{\Omega^2}\left(\frac{4\pi a_s}{m}\right)^2\sum_{\vec{pqr}}\left[n^0_{p\uparrow}n^0_{k\downarrow}-n^0_{q\uparrow}n^0_{r\downarrow}\right]\delta_{\vec{p}+\vec{k},\vec{q}+\vec{r}}\notag\\
&\times F^{(2)}(E_{pkqr},t) \label{eq:dn}
\end{align}
After setting  $g = 4\pi a_s/m$  in Eq.~\eqref{eq:Ek}   and subtracting $\Delta_{\text{kin}}(t)$, the divergence in the kinetic energy is cancelled and it is possible to obtain a finite result for the dynamics of the kinetic energy.  Furthermore,  for $p \gg k_F$ (compare Eq.~\eqref{eq:dn} and  Eq.~\eqref{eq:nplarge} after setting $g = 4\pi a_s/m$), we obtain:
\begin{equation}
\delta n_{p\gg k_F}(t) = \sum_{\sigma} n^{(2)}_{p\gg k_F,\sigma}\left(g = \frac{4\pi a_s}{m},t\right).  \label{eq:asymnp}
\end{equation}
This explains why adding $-\Delta_{\mathrm{kin}}(t)$ to the kinetic energy  cancels the  divergence which arises from the behavior of the instantaneous momentum distribution at large momenta. Indeed, 
this result provides the basis for our analysis of the dynamics of Tan's contact following the interaction quench, which is presented in Sec.~\ref{sec:contact}. In the following section, we shall evaluate the above expressions for a linear ramp of the interaction strength, for which $S(t)$ is given by Eq.~\eqref{eq:ramp_protocol}.
\section{Results for a linear ramp quench}\label{sec:ramp}

 In this section, we obtain explicit results for quenches  in which the interaction strength is ramped up linearly in time and 
 $S(t)$ is given by Eq.~\eqref{eq:ramp_protocol}.
As we have seen above, the first order correction to the total energy is just the expectation value of $V$ 
in the initial state (cf. Eq.~\ref{eq:U}) multiplied by $S(t)$. At second order in the scattering length $a_s$, using  
Eq.~\eqref{eq:Etot1} and evaluating the time integrals for the functions $F^{(1)}(t)$ (cf. Eq.~\ref{eq:f1}) and $F^{(2)}(t)$ (cf. Eq.~\ref{eq:f2}) yields  the following expression for the time dependence function,  $F_{\mathrm{tot}}(E,t) = F_{\mathrm{tot}}(E,T_r,t)$:
 \begin{align}
F_{\mathrm{tot}}(E,T_r,t) 
&=G(Et,E T_r) F_{\mathrm{tot}}^{\mathrm{eq}}(E),
\label{eq:Ftot}\\
G(x,y) &= \frac{  
 \theta(x-y) H(y)+\theta(y-x) H(x) }{(2 y)^2},\notag\\ 
H(x) &=  x^2-4\sin^2\left(\frac{x}{2}\right).
 \end{align}
For the purpose of  the discussion below, we have introduced a  scaling function,  $G(x,y)$, along with $F_{\mathrm{tot}}^{\text{eq}}(E)=E^{-1}$. The latter is the form that $F_{\mathrm{tot}}(E,t)$ takes in the adiabatic limit (see Appendix~\ref{sec:F}).
 
Let us next consider the  form of $F_{\mathrm{tot}}(E,T_r,t)$ in specific limiting cases. For a sudden quench,
$T_r \to 0$ and $G(Et,ET_r) = 0$. Therefore,
 \begin{align}
F^{\mathrm{sudden}}_{\mathrm{tot}}(E,t)=\lim_{T_r\to 0} F_{\mathrm{tot}}(E,T_r,t)=0,
 \end{align}
which implies that, to second order, the shift to total energy vanishes. Indeed, this  is expected from energy conservation in the limit of a sudden quench. To this this, let us compute the total energy directly for a sudden quench:
\begin{align}
E^{\mathrm{sudden}}_{\mathrm{tot}}(t>0)&=\langle e^{iHt}H(0^+) e^{-iHt}\rangle,\notag\\
&=\langle H(0^+) \rangle,\notag\\
&=E^{(0)}_{\text{tot}}  +\langle  V \rangle,\label{eq:enc}
 \end{align}
where $E^{(0)}_{\text{tot}}= \langle  H_0 \rangle$ is the  total energy before the quench (and $\langle \ldots \rangle$ is the average over the initial state). In other words, the dynamics of the total energy in the sudden quench limit simply amounts to a  constant energy shift happening at $t = 0$. The shift is the expectation value of the interaction in the initial state, which is first order in $a_s$. Hence,  not only the $O(a^2_s)$ term vanishes (in agreement with our findings) but so do all higher order corrections.

 However, in the adiabatic limit where $T_r\to +\infty$, we have
\begin{align}
F_{\mathrm{tot}}^{\mathrm{adiabatic}}(E,t)&=\lim_{T_r\to +\infty} F_{\mathrm{tot}}(E,T_r,t),\notag\\
&=\left[\theta(T_r-t)\frac{t^2}{T_r^2}+\theta(t-T_r)\right]F^{\mathrm{eq}}_{\mathrm{tot}}(E),\notag\\
&=S(t,T_r)^2F_{\text{tot}}^{\text{eq}}(E).\label{eq:Framp}
\end{align}
This means that at times $t < T_s$, there is a transient during which energy grows quadratically in time. The growth  saturates to the equilibrium value for $t >T_r$, i.e. the interaction strength reaches its full value. 

  We have numerically evaluated the momentum integrals and obtained the evolution  the second-order correction to the total energy and the instantaneous momentum distribution following a linear ramp in the interaction strength for a uniform Fermi gas in three dimensions.   The left panel of Fig.~\ref{fig:ramp}  shows the time evolution of second order correction to the total energy for different ramp times $T_r$. As  the sudden quench limit where $T_r\to 0$ is approached, the second order energy shift vanishes, as explained above.  In the opposite limit, as the ramp time $T_r$ increases, second-order energy shift saturates at a value that approaches the equilibrium  energy shift after initially growing quadratically at short times. 
 
  Concerning the dynamics of the instantaneous momentum distribution,  as discussed in Sec.~\ref{sec:md}, it is described by the function $F^{(2)}(E,T_r, t)$. For a linear ramp in the interaction strength,  this function can be written as follows (see Appendix~\ref{sec:F} for the details) :
\begin{equation}
F^{(2)}_{\text{ramp}}(E,T_r,t) =G^{(2)}(Et,ET_r)F^{(2)}_{\mathrm{eq}}(E),
\label{eq:F2}
\end{equation}
where 
\begin{align}
G^{(2)}(x,y) &= \frac{1}{x^2}\left\{2+\theta(y-x)\left[x^2-2\cos(x) \right. \right.  \notag\\
&\left. \left.  +2 x \sin(y-x)- 2x \sin(y) \right] \right.\notag\\
&\left.+\theta(x-y)\biggl[y^2-2\cos(y)-2y \sin(y)\biggr]\right\} \label{eq:gs2}	
\end{align}
is a scaling function of the dimensionless variables $x = ET_r$ and $y = E t$   and $F^{(2)}_{\text{eq}}(E)=1/E^2$ is the equilibrium value  of  $F^{(2)}(E,t)$. Thus,  as the sudden quench limit where $T_r \to 0$ is approached
\begin{multline}
F^{(2)}(E,T_r, t) = \left[ 4 \sin^2\left(\frac{E t}{2}\right)  +  E T_r \sin(E t) \right. \\ 
 \left.  + O(T^2_r)  \right] F^{(2)}_{\text{eq}}(E). \label{eq:gs2st}
\end{multline}
Therefore, we see that  the second order correction to instantaneous momentum distribution reaches  a stationary value that obeys the following relation:
\begin{equation}
 \lim_{t\to +\infty} \lim_{T_r\to 0} n^{(2)}_{k\sigma}(t,T_r)= n^{(2,\textrm{st})}_{k} = 2 n^{(2,\mathrm{eq})}_{k\sigma},
\end{equation}
where $n^{(2,\mathrm{eq})}_{k\sigma}$ is the second order correction to the equilibrium momentum distribution of the  Fermi gas with interaction strength equal to $g = 4\pi a_s/m$.  This result can be regarded as a generalization of the relationship obtained in Ref.~\cite{Moeckel2008,Moeckel2009} between the discontinuity of the momentum distribution at Fermi surface i.e. $Z(t) = \lim_{\delta\to 0} \left[ n_{k_F+\delta,\sigma}(t) - n_{k_F-\delta,\sigma}(t) \right]$  in the pre-thermalized regime
and in equilibrium:
\begin{equation}
1-Z_{\mathrm{st}}=2(1-Z_{\mathrm{eq}}),\label{eq:zst}
\end{equation} 
where $Z_{\mathrm{st}}$ and $Z_{\mathrm{eq}}$ stand for the stationary (i.e. pre-thermalized) and equilibrium values of $Z$, respectively.  

 On the other hand, in  the adiabatic limit where $E_F T_r \gg 1$, we have 
\begin{equation}
F^{(2)}(E,T_r, t) = \left[1  - \frac{2 \sin(E t)}{E T_r} + O(T^{-2}_r) \right] F^{(2)}_{\text{eq}}(E)
\end{equation}
Hence, the long time behavior of the instantaneous momentum distribution
approaches the equilibrium momentum distribution with an interaction strength $g = 4\pi a_s/m$,
i.e.
\begin{equation}
\lim_{t\to +\infty} \lim_{T_r \to +\infty} n^{(2)}_{k\sigma}(t,T_r)  = n^{(2,\textrm{eq})}_{k\sigma},
\end{equation}
as expected. 

 In the right-hand panel of Fig.~\ref{fig:ramp}, we have plotted the second-order correction to the instantaneous momentum distribution $n^{(2)}_{p,\sigma}(t)$ for $\sigma = \uparrow$. For reasons of experimental interest, we  consider an initial (non-interacting) state at finite temperature   $T = 0.1 T_F$, where $T_F = E_F = k^2_F/2m$ is the Fermi temperature (in $\hbar= k_B = 1$ units),  rather than the zero-temperature  ground state, as considered in previous studies~\cite{Moeckel2008,GGEpreth_2011_kollar,nessi_shorttime2014}.
Notice that the momentum distribution  reaches  a stationary value for $t/T_r\gg 1$. This
follows from the  asymptotic behavior of $F^{(2)}(E,T_r, t)$ for $t/T_r \gg 1$:
 \begin{equation}
 F^{(2)}(E,T_r,t\gg T_r) \simeq \left[ 1 + \frac{4 \sin^2(E T_r/2)}{E^2T^2_r} \right] F^{(2)}_{\textrm{eq}}(E). 
 \end{equation}
 Thus, for any finite  $T_r\neq 0$,  the long time limit of instantaneous momentum distribution for $k \sim k_F$  (see dashed line in Fig.~\ref{fig:ramp}) differs  from the equilibrium value $n^{(2,\textrm{eq})}_{k}$.

\section{Dynamics of the contact}\label{sec:contact}
\begin{figure}[b]
 \includegraphics[width=\columnwidth]{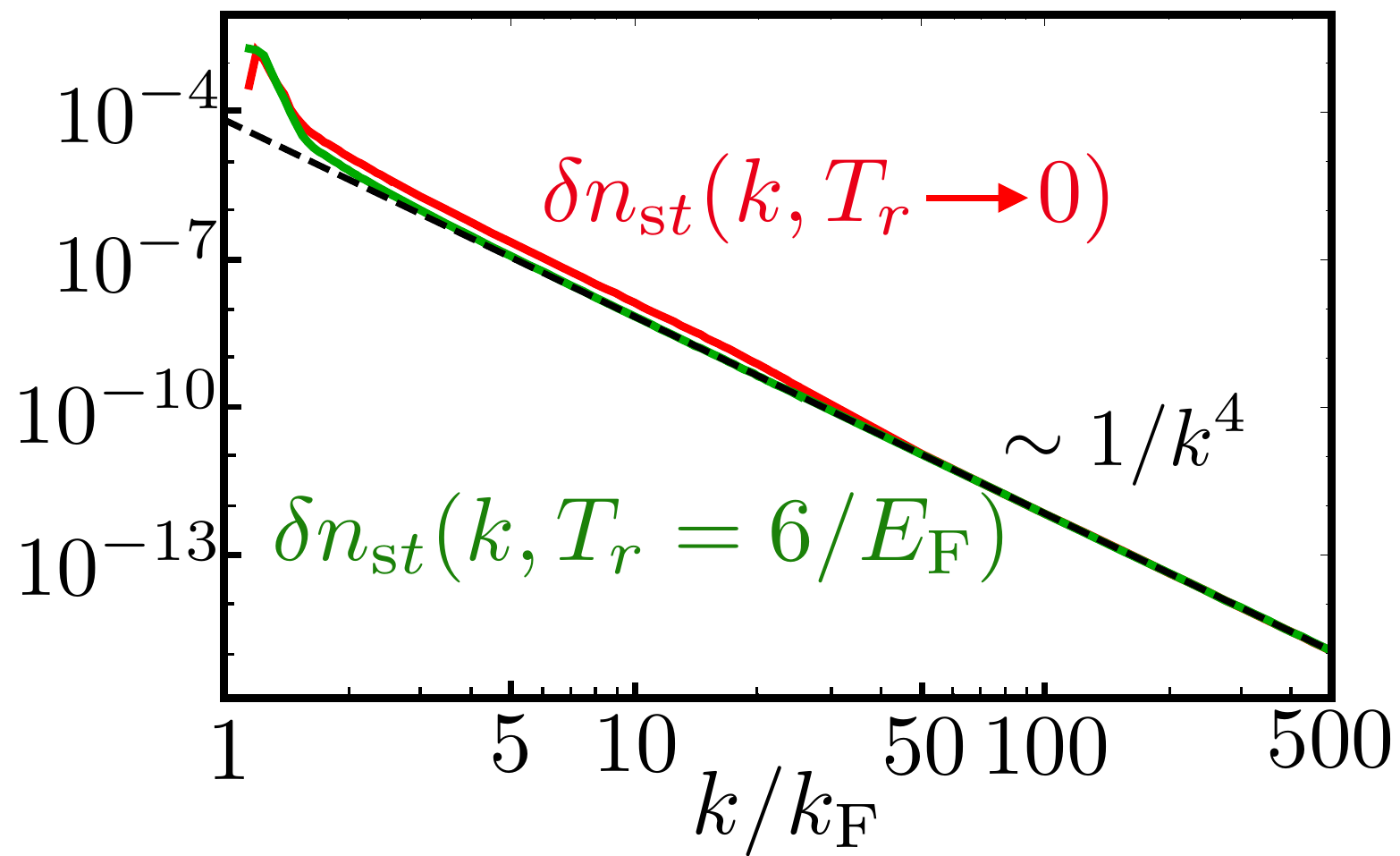}
 \caption{(color line) Log-log plot of the stationary value of the second order correction to the momentum distribution  $\delta n_{st}(k,T_r)$ derived from the renormalization procedure explained in Sec.~\ref{sec:re}, for $E_{\mathrm{F}}t=20$. The high-momentum   $\sim k^{-4}$ tail is displayed (dashed line).  Note that the  curves for the sudden quench limit $T_r\to 0$ and  $T_r E_F = 6$  exhibit the same asymptote for $k/k_F \gg 1$. Closer inspection reveals a crossover of the  $T_r E_F = 6$  curve between two different asymptotes. See explanation in Sec.~\ref{sec:inbtw}.  The interaction strength is chosen as $k_{\mathrm{F}} a_s=0.02$ and $T/T_F=0.1$.}\label{fig:asympt}
\end{figure}
 
In the previous section, we have studied the behavior of the perturbative corrections to the total energy and the instantaneous momentum distribution $n_{k\sigma}(t)$  for $k \sim k_F$ ($k_F$ being the Fermi momentum). In this section, we turn our attention to the asymptotic behavior of the latter for $k \gg k_F$. In this regime, $n_{k\sigma}(t)$ exhibits a $k^{-4}$ decay, similar to its behavior in equilibrium~\cite{shinatan_contact}. However, the pre-factor (the non-equilibrium version of Tan's contact) depends both on time, $t$ and the energy, $\epsilon_k = k^2/2m$.

  In connection with the asymptotic behavior of the instantaneous momentum distribution, in Sec.~\ref{sec:reneq} we have shown that the correction that  removes of the divergences from the perturbative expression for the total energy  factorizes into  a kinetic and an interaction-energy contribution.  The  correction to the kinetic  energy,  $\Delta_{\mathrm{kin}}(t)$ can be  written in terms of the function $\delta n_k(t)$  (see Eq.~\eqref{eq:renkin} and  following equations).  As  shown in Sec.~\ref{sec:reneq}, this function also yields the asymptotic behavior of  the  instantaneous momentum distribution,  $\sum_{\sigma}  n_{k\sigma}(t)$ for $k\gg p_F$ (see Eq.~\ref{eq:asymnp}). Hence, it is possible to extract the dynamics of Tan's contact by studying the asymptotic behavior of Eq.~\eqref{eq:dn}. To this end, let us rewrite $\delta n_k(t)$ as follows:
\begin{multline}
\delta n_k(T_r,t)=\frac{4}{\Omega^2}\left(\frac{4\pi a_s}{m}\right)^2\sum_{\vec{qr}}\left[n_{q\sigma}^0n^0_{r\alpha}-n^0_{k\sigma}n^0_{|\vec{q}+\vec{r}-\vec{k}|,\alpha}\right]\\
\times
F^{(2)}_{\text{ramp}}\left[\frac{k^2}{m}\left( 1+\frac{\vec{q}\cdot\vec{r}}{k^2}-\frac{\vec{k}\cdot(\vec{q}+\vec{r})}{k^2}\right),T_r,t\right],\label{eq:Cneq}
\end{multline}
where have implemented momentum conservation by requiring that $\vec{p} = \vec{q}+\vec{r}-\vec{k}$. In order to evaluate the above integrals numerically in the large $k$ limit, it is convenient to parametrize~\cite{Doggen:2015_contact}  $\vec{q}=(\vec{p}+\vec{k})/2+\vec{s}$, $\vec{r}=(\vec{p}+\vec{k})/2-\vec{s}$. The result of the numerical evaluation of $\delta n_k(T_r,t)$ is shown in Figure~\ref{fig:asympt} as a function of $k/k_F$  for $t = 20 E^{-1}_{\mathrm{F}}$. For this time,  $\delta n_k(T_r,t)$ has reached a stationary value. At large $k/k_F$,   we observe that it exhibits $1/k^4$ dependence. However, 
the detailed asymptotic behavior of $\delta n_k(T_r,t)$ is quite rich and also depends on  the ramp time $T_r$, as shown in Fig.~\ref{fig:asympt}. Indeed, a careful comparison the results for $T_r = 6 E^{-1}_{\mathrm{F}}$ and $T_r \to 0$ is worth here. For the former   value of $T_r$,  the asymptotic $\sim k^{-4}$ behavior is reached for  smaller values of $k/k_F$ than
for the $T_r\to 0$ results, which approach the sudden quench. In addition, close inspection of the curves shows that $\delta n_k(k,T_r \to 0)$ first approaches the $k^{-4}$ with a different coefficient, and only at large $k/k_F$ finally converges to the same asymptote as $\delta n_k(T_r = 6 E^{-1}_F,t)$. We  explain this behavior below. 

 Since the asymptotic behavior of the $\delta n_k(T_r,t)$ appears to be rather complex, a na\"ive generalization of  Tan's contact from the equilibrium case, i.e.
\begin{equation}	
C_{\mathrm{Tan}}(t)=\lim_{k\to \infty}k^4\delta n_{k}(t), \label{eq:ctrt}
\end{equation}
does not suffice to fully capture it. Thus,  in order to illustrate this point,  let us analyze the asymptotic behavior of the instantaneous momentum distribution in the sudden quench and adiabatic limits. 

\subsection{Sudden quench limit}
\begin{figure*}[t]
 \includegraphics[width=\textwidth]{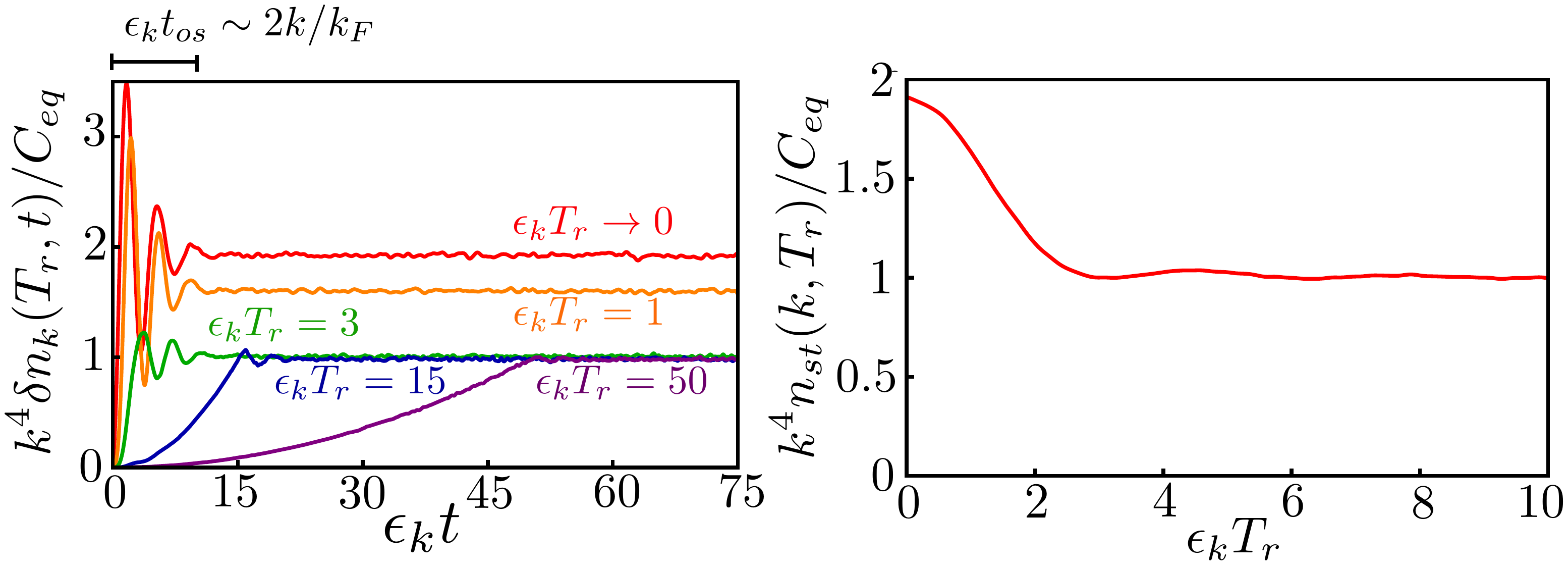}
 \caption{(color line) (Left panel) Coefficient controlling the asymptote of the $\sim k^{-4}$ tail of the instantaneous momentum distribution, i.e. $k^4\delta n_k(T_r,t)$, divided by the equilibrium value of Tan's contact, $C_{\mathrm{eq}}$. The calculation is carried out to second order in the scattering length $a_s$ for $k=5k_{\mathrm{F}}$  for a linear ramp of the interaction strength with characteristic time $T_r$. The time is measured in units of $\epsilon^{-1}_k$. For small $\epsilon_k T_r$,  after some oscillatory behavior,   the asymptote saturates to a stationary value which differs from $C_{\mathrm{eq}}$. As  $\epsilon_k T_r$ increases, the short time behavior exhibits a growth that is quadratic in time for $t < T_r$ and reaches the equilibrium value of Tan's contact. (Right panel) Stationary value of the asymptote $k^4\delta n_k(T_r,t)$ as a function of the  ramp time $T_r$. The dependence on $T_r$ shows a crossover between the sudden and adiabatic limits.  The temperature and interaction strength are chosen as in Fig.~\ref{fig:asympt}.}\label{fig:contact}
\end{figure*}
 In the sudden quench limit, using the following property (cf. Eqs.~\ref{eq:gs2} and Eq.~\ref{eq:gs2st}):
\begin{equation}
\lim_{ET_r\to0}G^{(2)}(Et,ET_r)=4\sin^2(Et/2),
\end{equation}
we obtain the following  behavior for the $k^{-4}$ asymptote:
\begin{align}
C^{\mathrm{sudden}}_{\mathrm{neq}}(t\gg1/\epsilon_k)&=\lim_{k\to\infty}\lim_{\epsilon_kT_r\to0}k^4\delta n(k,t\gg1/\epsilon_k),\notag\\
&=\frac{4}{\Omega^2}\left(4\pi a_s\right)^2\sum_{q,r}n_q^0n_r^0,\notag\\
&= 2C_{\mathrm{eq}},
\end{align}
where $C_{\mathrm{eq}}$ is the equilibrium contact calculated to leading order in perturbation theory (see Ref.~\cite{Doggen:2015_contact} and below). The above relation is similar to the relation obeyed in the pre-thermalized regime by the discontinuity of  the momentum distribution at the  Fermi momentum, Eq.~\eqref{eq:zst}. 

\subsection{Adiabatic limit}
In the adiabatic limit where $\epsilon_kT_r\to\infty$, we obtain the following expression for the asymptote:
\begin{align}
C_{\mathrm{neq}}^{\mathrm{adiabatic}}(t,T_r)&=\lim_{k\to\infty}\lim_{\epsilon_k T_r\to\infty} k^4\delta n_k(T_r,t),\notag\\
&=\left[\theta(T_r-t)\frac{t^2}{T_r^2}+\theta(t-T_r)\right]C_{\mathrm{eq}},\notag\\
&=\left[S(t,T_r)\right]^2 C_{\mathrm{eq}},\label{eq:Cramp}
\end{align}
Thus, like the total energy energy there is a transient  for $t <T_r$ where it grows quadratically in time. However, after the interaction reaches its full value at $t \ge T_r$, it saturates to the equilibrium value, $C_{\mathrm{eq}}$, in the steady state.

\subsection{In between}\label{sec:inbtw}

  Finally, let us consider  the general case of a finite ramp time $T_r > 0$. The left panel in Fig.~\ref{fig:contact} shows the time evolution of the asymptote of $k^4\delta n(k,T_r,t)$ normalized to the equilibrium contact for  different ramp times, $T_r$ (time is measured in $\epsilon^{-1}_k$ units). At short times, the asymptote of $k^4\delta n(k,T_r,t)$ oscillates with a period $t_{\mathrm{os}}\sim 2 \left(k/k_{F}\right) \epsilon^{-1}_k$. As $k\to\infty$ the oscillation dies out and, at large $\epsilon_k t$, the asymptote of  $\delta n_k(T_r,t)/C_{\mathrm{eq}}$ approaches a constant. The value of the latter varies between $1$ and $2$,  depending on the value of  $T_r$. The dependence on $T_r$ is shown on the right panel of Fig.~\ref{fig:contact}, which illustrates how the asymptotic behavior of $k^4 \delta n_k(T_r,t)$ crosses over from the adiabatic limit where
 \begin{equation} 
    k^4 \delta n_k(T_r,t)\to C_{\mathrm{eq}}
 \end{equation}   
to the sudden-quench limit where 
\begin{equation}
k^4 \delta n_k(T_r,t) \to 2 C_{\mathrm{eq}}.
\end{equation}
The crossover happens for $T_r \sim \epsilon^{-1}_k$. The explanation for this 
 behavior is as follows: For large but fixed $k\gg k_F$, any ramp of the interaction strength 
in a time much longer than $\epsilon^{-1}_k$ is regarded by the  fermions at  momentum $\vec{k}$ as adiabatic and therefore the behavior of asymptote approaches the adiabatic limit.  However, for $T_r \ll \epsilon^{-1}_k$, the fermions at momentum $\vec{k}$ experience the quench as sudden, and therefore the asymptote approaches the sudden limit. 

\section{Conclusions}\label{sec:conclu}

In conclusion, we have discussed  the renormalization of total energy obtained from the single-channel model in the context of interaction quenches. The resulting expressions are finite and, when evaluated for a linear ramp in the interaction strength starting from an non-interacting state in a two-component Fermi gas, allow us to study the crossover from the sudden-quench to the adiabatic limit. In addition, we have also studied the behavior of instantaneous momentum distribution. Thus, we found signatures of the pre-thermalization emerging at short to intermediate times after  a linear ramp of the interaction.  We have shown that the pre-thermalization signatures persist even at finite temperatures and they are also visible in the high-momentum tail of the distribution function. These results are important for the experimental characterization of this nonequilibirum state.   

 We have analyzed the dynamics of  the high momentum tail of the instantaneous momentum distribution, which is related to the non-equilibrium dynamics of the Tan's contact. Thus, we have uncovered an interesting crossover from adiabatic to sudden-quench dynamics in the asymptotic behavior of the momentum distribution as a function of the ramp time and the momentum scale of the fermions at the tail. Although explicit results were obtained only for a specific quench protocol that assumes a linear
ramp of the interaction strength, they should also hold for more general quenches provided the ramp time $T_r$ is replaced by the relevant switching-on time scale of the quench protocol.

 The results reported in this work can be  experimentally verified by preparing a three-dimensional two-component Fermi gas with an accessible broad Feshbach resonance in a non-interacting state, and quenching it to an interacting state. The total energy  dynamics can be accurately measured from the time-of-flight images obtained by  turning the scattering length to zero before releasing the gas from the trap.  In addition, the renormalization method described here can be readily applied to the computation of other interesting nonequilibrium properties such as the full work distribution. In addition, in future work it would be interesting to extending it beyond the second order in the scattering length.

\acknowledgements

We thank Y. Takahashi and Y. Takasu, and S.-Z. Zhang  for  useful discussions and commennts. We acknowledge support from the Ministry of Science and Technology of Taiwan through grants 102-2112-M-007-024-MY5 and 107-2112-M-007-021-MY5, as well as from the National Center for Theoretical Sciences (NCTS, Taiwan).

\appendix
  
\section{Time dependence of $F^{(1)}, F^{(2)}$ and $F_{\mathrm{tot}}$}\label{sec:F}

In the previous sections, we have used a number of results for the functions  $F^{(1,2)}(E,t)$, $F_{\mathrm{tot}}(E,t)$. In this section, we obtain their form  for a linear ramp as well as several other important limiting cases, namely the sudden-quench  and adiabatic limits. Let us recall that
\begin{align}
F_{\mathrm{tot}}(E,t)&=  S(t) F^{(1)}(E,t)-\frac{E}{2}F^{(2)}(E,t)\label{eq:Fapp},
\end{align}
describes the time dependence for the total energy, and
\begin{align}
F^{(1)}(E,t)&=\int\limits_{-\infty}^{t} \sin\left[E(t-t_1)\right] S(t) dt_1,\\
F^{(2)}(E,t)&=-\int\limits_{-\infty}^{t}dt_1 \int\limits_{-\infty}^{t} dt_2\:   S(t_1)S(t_2) e^{iE(t_2-t_1)}
\end{align}
describe the time dependence for interaction and kinetic  energy (momentum), respectively.  
For a ramp quench, setting $S(t,T_r)=\theta(t)[\theta(t-T_r)+\theta(T_r-t)t/T_r]$, we can derive:
\begin{align}
F^{(1)}_{\text{ramp}}(E&,T_r,t)=\frac{1}{E^2T_r}\biggl\{\theta(T_r-t)\biggl[Et-\sin(Et)\biggr]\\
&+\theta(t-T_r)\biggl[ET_r-\sin(Et)+\sin(E(t-T_r)\biggr]
\biggr\},\label{eq:Fr1}\\
F^{(2)}_{\text{ramp}}(E&,T_r,t)=\frac{1}{E^4T_r^2}\biggl\{2+\theta(t-T_r)\biggl[E^2T_r^2\notag\\
& -2\cos(ET_r) +2ET_r[\sin(E(t-T_r))-\sin(Et)]\biggr]\notag\\
&+\theta(T_r-t)\biggl[E^2t^2-2\cos(Et)-2Et \sin(Et)\biggr]\biggr\},
\label{eq:Fr2}
\end{align}
They describe the time dependence of total energy shift:
\begin{align}
 F_{\mathrm{tot}}(E,T_r,t)&=\frac{1}{2E^3 T_r^2}\biggl\{\theta(t-T_r)\biggl[E^2T_r^2-\sin^2(ET_r/2)\biggr] \notag\\
& \qquad+\theta(T_r-t)\biggl[E^2t^2-4\sin^2(Et/2)\biggr]\biggr\}\label{eq:Ftotapp}.
 \end{align}
For the the sudden quench $S(t)=\theta(t)$. We found the two time dependent functions $F^{(1)}(E,t)$ and $F^{(2)}(E,t)$:
\begin{align}
F^{(1)}_{\text{sudden}}(E,t)&=\frac{2\sin^2\left(Et/2\right)}{E},\label{eq:FS1}\\
F^{(2)}_{\text{sudden}}(E,t)&=\frac{4\sin^2\left(Et/2\right)}{E^2},\label{eq:FS2}
\end{align}
from which, the function that controls the  time dependence of total energy in the sudden quench limit:
\begin{align}
F_{\mathrm{tot}}^{\mathrm{sudden}}(E,t>0)&=  F^{(1)}(E,t)-\frac{E}{2}F^{(2)}(E,t) \notag\\
		&=0,\label{eq:zero}
\end{align}
as required by the conservation of total energy (see discussion in Sec.~\ref{sec:ramp}, Eq.~\ref{eq:enc}). In the equilibrium limit, using $S(t)=e^{-\eta |t|}$  with $\eta\to 0^+$ to describe the adiabatic switching of the interaction, we  obtain:
\begin{align}
F^{(1)}_{\mathrm{eq}}(E,t)&=\frac{1}{E},\label{eq:FS1a}\\
F^{(2)}_{\mathrm{eq}}(E,t)&=\frac{1}{E^2},\label{eq:FS2a}
\end{align}
which lead to the equilibrium results. Hence,
\begin{align}
F^{\mathrm{eq}}_{\mathrm{tot}}\left(E,t\right)&=\frac{1}{2E}. \label{eq:ftot_adiabatic}
\end{align}
\bibliography{cite}

\end{document}